\input{aipcheck}
\documentclass[final]{aipproc}
\layoutstyle{6x9}

\begin{document}

\title{Quantum Optical Heating in Sonoluminescence Experiments}


\author{Andreas Kurcz}{
  address={School of Physics and Astronomy, University of Leeds, Leeds, LS2
    9JT, United Kingdom}
}

\author{Antonio Capolupo}{
  address={School of Physics and Astronomy, University of Leeds, Leeds, LS2
    9JT, United Kingdom}
}

\author{Almut Beige}{
  address={School of Physics and Astronomy, University of Leeds, Leeds, LS2
    9JT, United Kingdom}
}

\begin{abstract}
Sonoluminescence occurs when tiny bubbles filled with noble gas atoms are
driven by a sound wave. Each cycle of the driving field is accompanied by a
collapse phase in which the bubble radius decreases rapidly until a short but
very strong light flash is emitted. The spectrum of the light corresponds to
very high temperatures and hints at the presence of a hot plasma core. While
everyone accepts that the effect is real, the main energy focussing mechanism is highly controversial. Here we suggest that the heating of the bubble might be due to a weak but highly inhomogeneous electric field as it occurs during rapid bubble deformations [A. Kurcz {\em et al.} (submitted)]. It is shown that such a field couples the quantised motion of the atoms to their electronic states, thereby resulting in very high heating rates.
\end{abstract}

\maketitle

%%%%%%%%%%%%%%%%%%%%%%%%%%%%%%%%%%%%%%%%%%%%
%% MAINMATTER
%%%%%%%%%%%%%%%%%%%%%%%%%%%%%%%%%%%%%%%%%%%%

\section{Introduction}

Sonoluminescence is a phenomenon that derives from the acoustic cavitation of
noble gas atoms \cite{Frenzel}. There are two classes of sonoluminescence: multi-bubble \cite{Walton,Mcnamara} and single-bubble sonoluminescence  \cite{Gaitan,Brenner}. Single bubble sonoluminescence is characterized by the emission of a strong light flash over a very short period of time from a single extremely hot gas bubble.
Under appropriate conditions, the acoustic force on a bubble can balance against its buoyancy, holding a bubble stable in the liquid by acoustic levitation. Such a bubble is typically quite small compared to an acoustic wavelength and is capable to confine the particles of the trapped van der Waals gas close to their covolume. For specialized conditions, a single, stable, oscillating gas bubble can be forced into such large amplitude pulsations that it produces sonoluminescence during each and every acoustic
cycle. 

\begin{figure}
\includegraphics[scale=0.4]{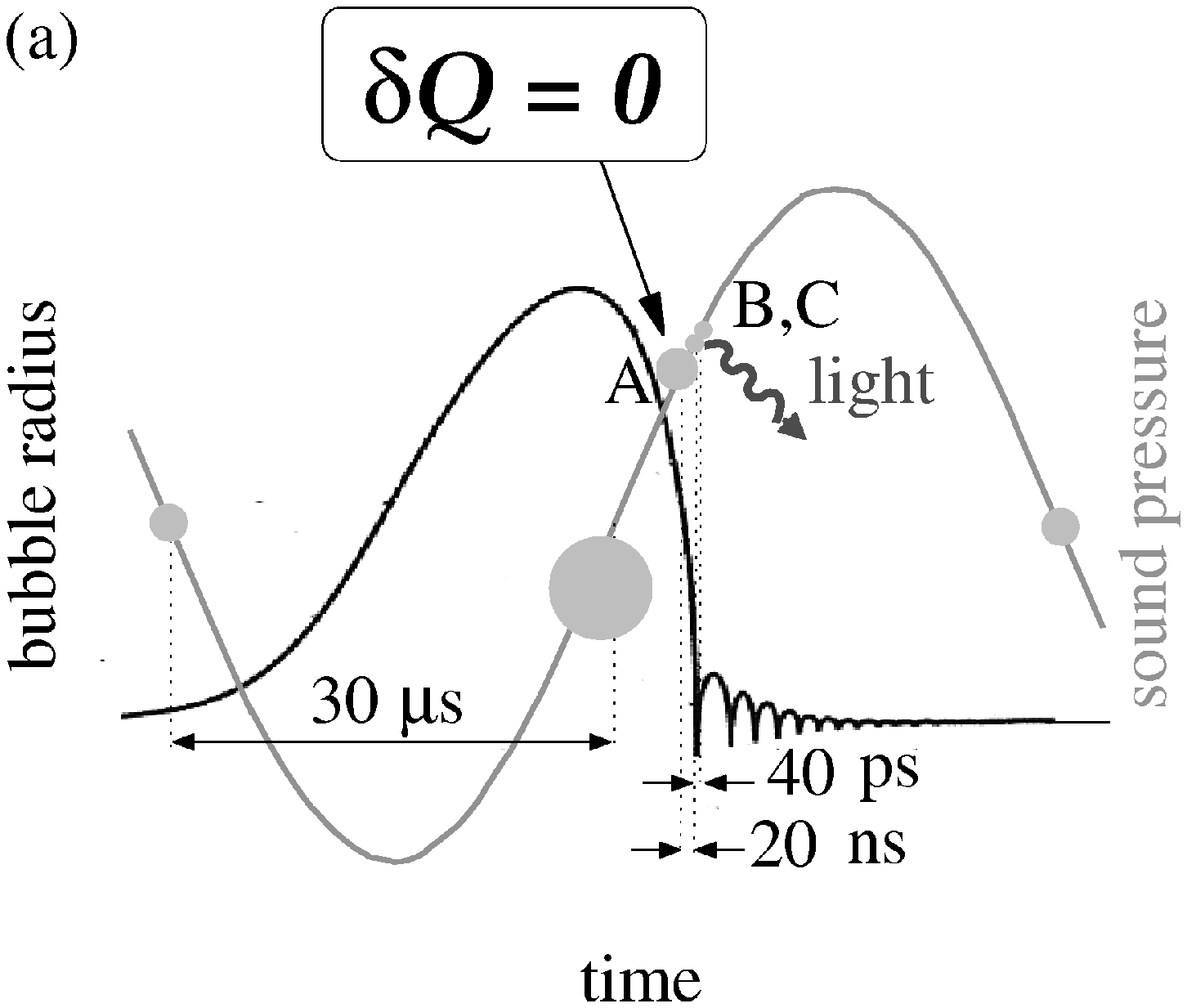}
\hspace{2cm}
\includegraphics[scale=0.4]{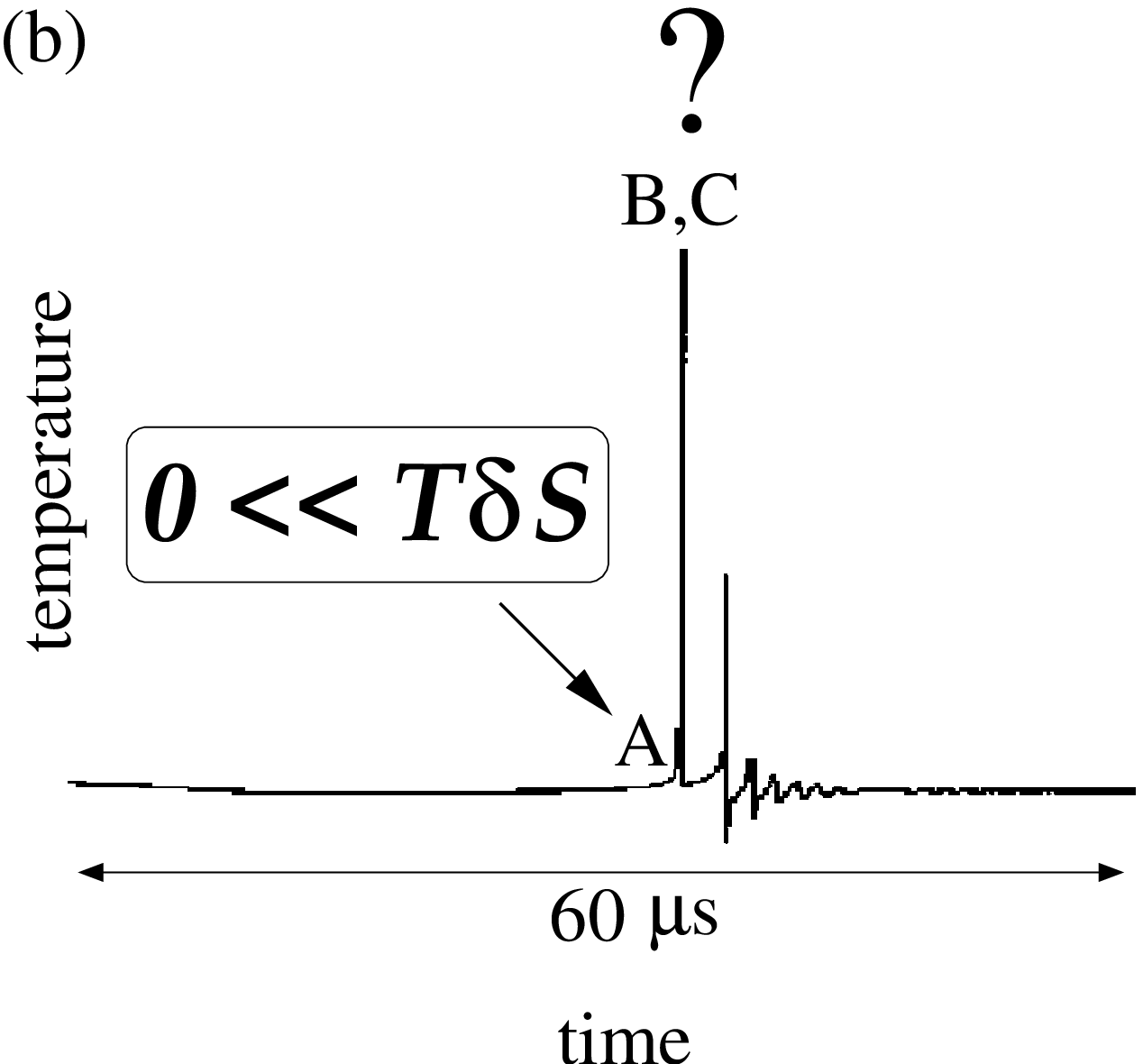}
\caption{A typical single-bubble sonoluminescence cycle. (a):
Time dependence of the driving sound pressure and the corresponding bubble
radius. (b): Time dependence of the temperature.}
\end{figure}

A typical single-bubble sonoluminescence cycle is shown in Fig. 1(a). Most of the
cycle, the bubble behaves isothermal (c.f.~Fig. 1(b)). Point $A$ marks the beginning of the collapse phase in which the bubble approaches its minimum radius of about $0.5 \, \mu $m very rapidly with supersonic speed. Here, the bubble becomes
thermically isolated from the surrounding liquid. In Point $B$, the temperature within the
bubble significantly increases with a heating rate of $10^{10} -10^{11}\,$K/s and a strong light flash emerges which last for about $40 \,$ps. Point C denotes the beginning of the expansion phase in which the bubble oscillates around its equilibrium radius until it
regains stability.

The emitted light mainly consist of a continuum 
of blackbody or Bremsstrahlung radiation. 
Detailed measurements of the light spectra indicate temperatures above 
$10^4 \, $K \cite{10h41,10h42,10h43}. It is even possible to 
observe light emission in the ultraviolet regime which hints at temperatures
of about $10^6\, $K in a bubble driven at $1\, $Mhz \cite{1Mhz}. Emission lines from transitions between high energy 
states of noble gas atoms which cannot be populated thermally
\cite{PlasmaSbsl1,PlasmaSbsl2} point at the formation of an {\em opaque} plasma core
\cite{Suslick}. Evidence for a plasma core has also been found in multi-bubble
sonoluminescence experiments \cite{PlasmaMbsl}.

The time dependence of the bubble radius and its nearly adiabatic compression
are theoretically well understood up to a certain point when it approaches the
minimum radius \cite{Brenner, Suslick}. What is the state of the bubble during the last part of the collapse phase and the conditions that lead to these enormous heating
rates up to very high temperatures is still controversial.  Here we summarise an idea which suggests that the heating is due to the presence of a highly inhomogenous electric field as it occurs during rapid bubble deformations \cite{Kurcz}. This field couples the motion of the noble atoms to their electronic degrees of freedom. When combined with spontaneous emission from the atoms, a quantum optical heating process can occur. Similar couplings are responsible for the cooling of ions in ion trap experiments \cite{Wineland}.

\section{The basic idea}
 
Approaching its minimum radius close to point $B$ in Fig. 1(a), the bubble is no longer in a thermal equilibrium. Suddenly, an increase in entropy occurs which is based on highly irreversible processes. It causes a temperature increase much higher than what can be caused by thermodynamic heating processes (c.f. Fig. 1(b)). In the following we address two questions: Why does the bubbles need to be filled with noble gas atoms? What is the main energy focussing mechanism during the collapse phase of the bubble?

Close to point $B$, the mean distance between the noble gas atoms becomes so small that interactions between them can be described by a Lennard-Jones
potential. Indeed, the physical condition of the bubble becomes that of a
solid state system. The atoms experience an equilibrium between repulsive
interatomic forces due to overlapping orbitals and attractive forces due to
the van der Waals interaction. Thus, any significant gain in temperature has
to be caused by vibrational motion driven into the quantum
regime. Furthermore, the presence of light requires the assumption of an {\em
open} quantum system. 

To model the resulting strong confinement of the atoms, we
place each of them into an approximately harmonic trapping potential. This
allows us to quantise the atomic motion during the collapse phase, just before
the maximum compression of the bubble. Around this point, the motional states
of each atom can be described by phonons with frequency $\nu$.
In the next section, we show that the gradient of an electric field inside
the bubble establishes a coupling between the electronic and the quantised
motional states of each noble gas atom. The origin of the field can be
explained by  an inhomogeneous charge distribution of ionized species from the dissolved liquid due to rapid bubble deformations.

For simplicity we assume that the
atoms are effective two-level systems with ground state $|0 \rangle$ and
excited state $|1 \rangle$. The corresponding interaction Hamiltonian contains
terms that result in the excitation and de-excitation of each atom accompanied
by the creation and the annihilation of a phonon. Also crucial is the presence
of a large spontaneous decay rate $\Gamma$ of the excited state $|1 \rangle$ which
keeps the atoms predominantly in their ground state. Although these processes
are highly non-resonant, they result in a significant change of the mean
phonon number per atom and increase the temperature inside the bubble by many
orders of magnitude, even within a few nanoseconds. 

Suppose an atom is initially in its ground state and possesses exactly $m$
phonons, as shown in Fig. 2a. We denote this state by $|0,m
\rangle$. Notice that phonons are bosons which are described by annihilation
operators $b$ with $[b,b^\dagger ] = 1$.
Consequently, a transition into the state $|1, m+1 \rangle$ occurs with a rate
proportional to $\sqrt{m+1}$, while the rate for a transition into the state
$|1, m-1 \rangle$ scales only as $\sqrt{m}$. Since the spontaneous decay rate
of the atom is relatively large, such a transition is immediately
followed by an irreversible and predominantly non-radiative transition back
into $|0 \rangle$. This transfers the atom either into its initial state $|0,m
\rangle$ or into the states $|0, m-1 \rangle$ and $|0, m+1 \rangle$,
respectively. The net effect is an increase of the mean phonon number per
atom, i.e.~heating, since the phonon population in the latter state is higher
than the phonon population in $|0, m-1 \rangle$.

\section{The time evolution of the system}

\begin{figure}
\includegraphics[scale=0.5]{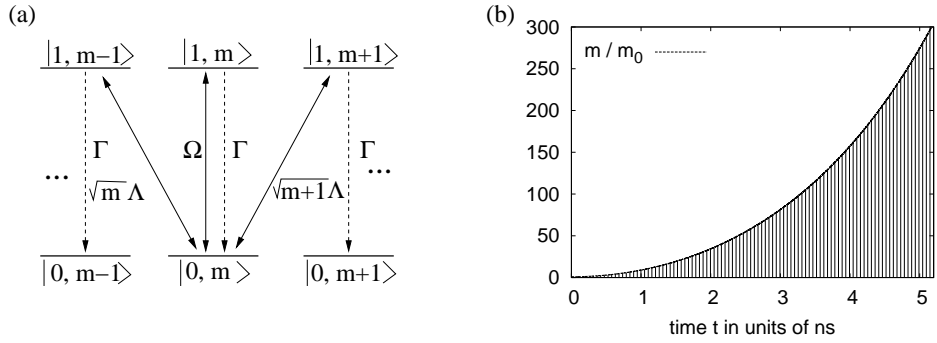}
\caption{(a): Level configuration of a single atom-phonon system indicating
  the immediately relevant transitions, if the atom is initially in $|0,m
  \rangle$. $\Omega$ and $\Lambda$ denote coupling constants and $\Gamma$ is
  the spontaneous decay rate of level 1.~(b): The mean phonon number $m$ as a
  function of time for $\nu = 10\,$MHz while $\Omega = 10^{6} \, $Hz, $\Lambda
  = 10^{12} \, $Hz, $\Gamma = 10^{13} \, $Hz $^1$, and $\omega_0 = 10^{15} \, $Hz. Good agreement is found between the numerical solution of the full rate equations (\ref{fullset}) and (\ref{fullset2}) and Eq.~(\ref{fullsolution}) (shaded area).}
\end{figure}

We now consider a {\em single} noble gas atom at the position ${\bf r}$. This atom is typical for the many atoms inside the bubble. Its dipole Hamiltonian equals
\begin{equation} \label{Hs}
H _{\rm int} = e\, {\bf D} \cdot {\bf E} ({\bf r})  
\end{equation}
with $e$ being the charge of a single electron, the (real) atomic dipole moment 
\begin{eqnarray}
{\bf D} =  {\bf D}_{01} \, \sigma^{-} + {\rm H.c.} \, , 
\end{eqnarray}
$\sigma^+ \equiv |1 \rangle \langle 0|$, $\sigma^- \equiv |0 \rangle \langle 1|$, and where ${\bf E}$ is the electric field inside the bubble. For simplicity, we assume that all field components point in the direction of a single unit vector $\hat {\bf k}$. This allows us to write ${\bf E} ({\bf r})$ as 
\begin{eqnarray}
{\bf E} ({\bf r}) = \sum_k {\bf E}_k \, {\rm e}^{{\rm i} k \hat {\bf k} \cdot {\bf r}} + {\rm c.c.}
\end{eqnarray}
with amplitudes ${\bf E}_k$ and wave vectors ${\bf k}= k \hat {\bf k}$. Moreover, we consider the atomic motion in the $\hat {\bf k}$-direction as quantised with $b$ being the corresponding phonon annihilation operator. Then $\hat {\bf k} \cdot ({\bf r} - {\bf R}) = \Delta x \big( b  + b  ^\dagger \big)$. Here ${\bf R} $ is the current equilibrium position of the noble gas atom with mass $M$ and $\Delta x=\sqrt{\hbar / 2 M \nu}$ is the width of its ground state wave function in the respective vibrational mode. If the atom is well localized within the wavelength of its trapping potential, the Lamb-Dicke approximation allows us to assume that
$\exp( {\rm i} k \hat {\bf k} \cdot ({\bf r}-{\bf R})) = 1 + {\rm i} k \Delta
x \big(b + b^\dagger \big)$ \cite{Wineland, LambDicke}. Substituting this into Eq.~(\ref{Hs}), we obtain the interaction Hamiltonian 
\begin{equation} \label{Hint}
H _{\rm int} = \hbar \Omega \, ( \sigma ^- + \sigma^+) + \hbar \Lambda  \, (b  + b ^\dagger) ( \sigma ^- + \sigma^+) 
\end{equation}
with the (real and positive) coupling constants 
\begin{eqnarray}
\Omega &\equiv & \left ( 2 e / \hbar \right ) \sum _k {\bf D}_{01} \cdot {\rm{Re}} \left (
  {\bf E}_k {\rm e}^{{\rm{i}} k \hat {\bf k} \cdot \bf{R}} \right ) \, , \nonumber \\
\Lambda &\equiv & - \left( 2 e ~ \Delta x / \hbar \right) \sum _k k  \, {\bf D}_{01} \cdot
{\rm{Im}} \left ( {\bf E}_k {\rm e}^{{\rm{i}} k \hat {\bf k} \cdot {\bf R}}
\right ) \, .
\end{eqnarray}
This Hamiltonian is essentially a Jaynes-Cummings Hamiltonian with $\Lambda$
being proportional to the gradient of $\Omega$ in the direction of the
quantised motion of the atom, i.e.~$\Lambda = \Delta x \, \hat{\bf k} \cdot
\nabla \Omega ({\bf R}) $. A strong atom-phonon coupling therefore does not
necessarily require the presence of a strong electric field. It only requires
a highly inhomogeneous field inside the bubble. In the following, we neglect
interactions between the noble gas atoms other than the ones already included
in the harmonic trapping potential of each particle. Dissipation in form
of spontaneous photon emission from the atomic state $|1 \rangle$ is 
taken into account by the master equation \cite{LambDicke}
\begin{equation} \label{master2}
\dot \rho = - \frac{{\rm i}}{\hbar} \left [ H _{\rm int} + \hbar \omega_0 \, \sigma ^+ \sigma ^- + \hbar \nu  \, b ^\dagger b  \, ,  \rho \right ]
+ \Gamma \left[ \, \sigma ^- \, \rho \, \sigma ^+ - \frac{1}{2} \sigma ^+ \sigma ^- \, \rho -  \frac{1}{2} \, \rho \,
\sigma ^+ \sigma ^- \, \right] \, .
\end{equation}
Here $\hbar \omega_0$ and $\hbar \nu$ are the energy of the atomic state $|1
\rangle$ and of a single phonon. 

Eq.~(\ref{master2}) can now be used to obtain
a closed set of rate equations. Its major quantities are the phonon number $m
\equiv \langle b ^\dagger  b  \rangle$ and
\begin{eqnarray} \label{coh}
X_{1,2} \equiv \langle
\sigma_{1,2} \rangle \, , ~~ X_3 \equiv \langle \sigma^+ \sigma^- - \sigma^-
\sigma^+ \rangle \, , ~~ Y_1 \equiv \langle b + b^\dagger \rangle \, , ~~ Y_2
\equiv {\rm i} \langle b - b^\dagger \rangle \, , \nonumber \\
Y_3 \equiv \langle b^2 + b^{\dagger \, 2} \rangle \, , ~~ Y_4 \equiv {\rm i} \langle b^2 -
b^{\dagger \, 2} \rangle \, , ~~  Z_{1,2} \equiv \langle \sigma_{1,2} ( b +
b^\dagger ) \rangle \, , ~~ Z_{3,4} \equiv {\rm i} \langle \sigma_{1,2} ( b -
b^\dagger ) \rangle
\end{eqnarray} 
with the Pauli operators $\sigma_1 \equiv \sigma^+ + \sigma^-$ and $\sigma_2 \equiv {\rm i} (\sigma^- - \sigma^+)$. Here we assume 
\begin{eqnarray} \label{paris}
\omega_0 \, \gg \, \nu, \, \Gamma, \, \Omega, \, \Lambda
~~ {\rm and} ~~ m \gg 1 
\end{eqnarray}
and approximate the expectation value of operators of the form $\langle B \sigma _3 \rangle$ by $\langle B \, \rangle \langle \sigma _{3} \rangle$. The latter applies when the expectation value of $B$ is about the same for an atom in $|0 \rangle$ and for an atom in $|1 \rangle$. Eq.~(\ref{master2}) then yields
\begin{eqnarray} \label{fullset}
&& \dot m = \Lambda  Z_3 \, , ~~  \dot X_3 = 2 (\Omega X_2 + \Lambda Z_2 ) -
\Gamma \left ( X_3 + 1 \right ) \, , ~~\dot Y_1 = - \nu Y_2  \, , ~~ \nonumber \\
&& \dot Y_2 = 2 \Lambda X_1 + \nu Y_1 \, , ~~ \dot Y_3 = - 2 (\nu
Y_4 + \Lambda Z_3) \, , ~~ \dot Y_4 = 2 (\nu Y_3 + \Lambda Z_1) \, ,
\end{eqnarray}
and
\begin{eqnarray} \label{fullset2}
&& \dot X_1 = - \omega_0 X_2 \, , ~~ \dot X_2 = - 2 (\Omega + \Lambda Y_1) X_3
+ \omega_0 X_1 \, , ~~ \dot Z_1 = - \omega_0 Z_2 \, , ~~ \dot Z_3 =  2 \Lambda - \omega_0 Z_4 \, , ~~~ \nonumber \\
&&  \dot Z_2 = - 2 (\Omega Y_1 +
\Lambda Y_3 +2 \Lambda m) X_3 + \omega_0 Z_1 \, , ~~ \dot Z_4 = - 2 (\Omega Y_2 + \Lambda Y_4 ) X_3 + \omega_0 Z_3 
\end{eqnarray}
up to first order in $1/\omega_0$. 
In the beginning of each sonoluminescence cycle the particles experience neither a strong trapping potential nor the presence of an inhomogeneous electric field inside the bubble. We can therefore assume that the coherences defined in Eq.~(\ref{coh}) are initially zero and that the atom is in its ground state. Condition (\ref{paris}) allows us to 
simplify the above rate equations via an adiabatic elimination of Eq.~(\ref{fullset2}). Doing so we obtain a set of equations where the derivatives of $X_3$, $Y_1$, and $Y_2$ decouple from the rest. Solving them for the case of a relatively strong atom-phonon coupling constant $\Lambda$ with 
$\Lambda \gg \Omega ~~ {\rm and} ~~ 4 \Lambda^2 \, > \, \nu \omega_0$ yields 
\begin{equation} \label{X3}
X_3 (t) = -1 \, , ~~  Y_1 (t) = \frac{4 \nu \Omega \Lambda}{\lambda^2 \omega_0} \cdot \big[ \cosh ( \lambda t ) -1 \big] \, , ~~ 
Y_2 (t) = -  \frac{4 \Omega \Lambda}{\lambda \omega_0} \cdot \sinh ( \lambda t)  
\end{equation}
with $ \lambda \equiv  \nu \left( 4 \Lambda^2 / \nu \omega _0 - 1 \right)^{1/2}$
up to first order in $1/\omega_0$. For times $t$ of the order of $1/\lambda$, $Y_1$ and $Y_2$ are of the order of $1/\omega_0$. Taking this into account, we find that $Z_1 = - 2 \Lambda (2 m + Y_3)/\omega_0$ and $Z_3 =  - 2 \Lambda Y_4/\omega_0$ in first order in $1/\omega_0$. The variables $m$, $Y_3$, and $Y_4$ in Eq.~(\ref{fullset}) hence evolve according to  
\begin{equation}\label{set}
 \dot m =  - {2 \Lambda^2 \over \omega_0} \, Y_4 \, , ~~ 
\dot Y_3 = {2 (2 \Lambda^2 - \nu \omega_0) \over \omega_0} \, Y_4 \, , ~~
\dot Y_4 =  - {8 \Lambda^2 \over \omega_0} \, m - {2 (2 \Lambda^2 - \nu \omega_0) \over \omega_0} \, Y_3 \, .
\end{equation}
For $m(0)=m_0$ and $Y_3(0)=Y_4(0)=0$, this yields
\begin{eqnarray} \label{fullsolution}
m(t) &=& m_0 + {8 \Lambda^4 \over \lambda^2 \omega _0^2} \cdot m_0 \, \sinh^2 ( \lambda t ) \, .
\end{eqnarray}
As one can see in Fig. 2(b), Eq. (\ref{fullsolution}) describes an approximately exponential
heating process as long as a relatively large decay $\Gamma$ secures that the
atom remains in the ground state predominantly 
\footnote{The assumption of a
  relatively high spontaneous decay rate can be justified by the presence of
  collective effects inside the van der Waals gas formed by the noble gas
  atoms.}. Taking into account
typical experimental parameters, the phonon energy in the bubble can easily
increase by a factor ten or more, even within a few nanoseconds. Using the
relation $m \cdot \hbar \nu = k_B T$, our model can easily predict
temperatures well above $10 ^4 \,$K inside the bubble.

\section{Conclusion}

We attribute the sudden concentration of energy in sonoluminescence
experiments to the heating of strongly confined noble gas atoms by a highly
inhomogeneous electric field. The time evolution of each atom is dominated by
non-energy conserving processes, which result in a permanent increase of its
mean phonon number $m$ when combined with spontaneous emission.
Our model does not contradict current models for the description of
sonoluminescence experiments, but explains previously controversial aspects of
this phenomenon. It is based on a quantum optical approach that is routinely
used to describe the laser cooling of tightly trapped ions \cite{Wineland}.

\begin{theacknowledgments}
  A. B. acknowledges a James Ellis University Research Fellowship from the Royal Society and the GCHQ. This work was moreover supported in part by the EU Research and Training Network EMALI and the UK Research Council EPSRC. 
\end{theacknowledgments}



\bibliographystyle{aipproc}   % if natbib is available

\bibliography{sample}

\IfFileExists{\jobname.bbl}{}
 {\typeout{}
  \typeout{******************************************}
  \typeout{** Please run "bibtex \jobname" to optain}
  \typeout{** the bibliography and then re-run LaTeX}
  \typeout{** twice to fix the references!}
  \typeout{******************************************}
  \typeout{}
 }

%%%%%%%%%%%%%%%%%%%%%%%%%%%%%%%%%%%%%%%%%%%
%% The following lines show an example how to produce a bibliography
%% without the help of the BibTeX program. This could be used instead
%% of the above.
%%%%%%%%%%%%%%%%%%%%%%%%%%%%%%%%%%%%%%%%%%%

\end{document}